\documentclass[preprint,12pt]{aastex}
\usepackage{natbib}

\newcommand{\Msun}{M_\sun}
\newcommand{\Mdot}{\dot{M}}
\newcommand{\fprime}{f'}

\begin{document}
\bibliographystyle{apj}
\title{Wolf-Rayet Mass-Loss Limits Due to Frequency Redistribution}
\author{Andrew J. Onifer}
\affil{Los Alamos National Laboratory, X-2, MS-B227, Los Alamos, NM 87545}
\email{aonifer@lanl.gov}
\and \author{Kenneth G. Gayley}
\affil{Department of Physics and Astronomy, University of Iowa, Iowa
City, IA 52242}

\begin{abstract}
The hypothesis that CAK-type line driving is responsible for the large
observed Wolf-Rayet (W-R) mass-loss rates has been called into
question in recent theoretical studies.  The purpose of this paper is
to reconsider the plausibility of line driving of W-R winds within the
standard approach using the Sobolev approximation while advancing the
conceptual understanding of this topic.  Due to the multiple
scattering required in this context, of particular importance is the
role of photon frequency redistribution into spectral gaps, which in
the extreme limit yields the statistical Sobolev-Rosseland (SSR) mean
approximation. Interesting limits to constrain are the extremes of no
frequency redistribution, wherein the small radii and corresponding
high W-R surface temperature induces up to twice the mass-loss rate
relative to cooler stars, and the SSR limit, whereby the reduced
efficiency of the driving drops the mass flux by as much as an order
of magnitude whenever there exist significant gaps in the spectral
line distribution.  To see how this efficiency drop might be
sufficiently avoided to permit high W-R mass loss, we explore the
suggestion that ionization stratification may serve to fill the gaps
globally over the wind.  We find that global ionization changes can
only fill the gaps sufficiently to cause about a 25\% increase in the
mass-loss rate over the local SSR limit.  Higher temperatures and more
ionization states (especially of iron) may be needed to achieve
optically thick W-R winds, unless strong clumping corrections
eliminate the need for such winds.
\end{abstract}


\section{Introduction}
\label{sec:intro}
Wolf-Rayet (W-R) star mass-loss rates are inferred to be as high as
several times $10^{-5} M_\odot \ yr^{-1}$, or $10^9$ times the mass-loss
rate of the Sun 
\citep{morris-etal2000,
  nugis-lamers2000}.  More importantly, their mass-loss rates are
significantly enhanced over their O-star progenitors, even though
their luminosity is similar.  \citet*[][hereafter CAK]{cak} showed
that O and B star winds can be driven by radiation pressure via the
opacity of a large number of UV spectral lines, so the question arises
if W-R winds may also be driven by line opacity
\citep[e.g.,][]{barlowe-etal1981, cassinelli-vanderhucht1987,
  lucy-abbott1993, springmann-puls1995}. Note this would require an
enhancement in the effective line opacity in a W-R star, and in
considering how this added opacity may come about, it is the goal of
this paper to develop a conceptual language for discussing what
additional difficulties emerge.

Excellent agreement with observations is being achieved by detailed
models \citep{hillier-miller1999, demarco-etal2000,
  crowther-etal2002}, but conceptual uncertainties remain, including
the maximum attainable driving efficiency, and the key differences
between O and W-R winds that generate higher mass loss.  Also, it is
not clear how sensitively the models rely on details in the line lists,
and it would be advantageous to be able to generate simplified models
that retain the key physics but are able to be extended to
computationally demanding investigations such as hydrodynamic
simulations. As a complement to such conceptually
complex analyses, we will isolate particular wind properties important
for driving W-R winds using simplified analytic expressions and approximations.

Because we are concerned with the mass-loss rate and not the global momentum 
deposition (the latter stems in part from the
former), our analysis is restricted to a local model at the critical point,
which is the point deep in the wind where the mass flux is most difficult to drive.
In addition, for simplicity and to get results that are as general as possible,
we avoid identifying model-specific parameters
such as the velocity law 
and the location where the critical point appears, as these are not fundamental to
the overall mass driving.  The goal is to analyze the impact of
frequency redistribution on the mass flux in any optically
thick line-driven flow, rather than to
model a specific W-R star or create a grid of such models.

Since the role of line opacity is the central issue, and because
the flows are observed to be highly supersonic, considerable
conceptual progress may be made by invoking the Sobolev approximation,
as is typical in OB-star applications.
Some authors 
have questioned the use of the Sobolev
approximation in optically thick flows such as Wolf-Rayet winds.  For
example \citet{grafener-hamann2005} employ models in which the CAK $\alpha$
parameter, which describes the ratio of optically thick to optically
thin lines, is essentially 0, presumably due to the inclusion of very
large turbulent velocities.  \citet{nugis-lamers2002} argue that the
wind mass-loss rate is determined near the sonic point, and therefore use
static Rosseland mean opacities in their analysis (though they use
CAK-type line driving in the wind beyond the sonic point).  In
fact the Sobolev approximation only breaks down if optically {\it thick}
lines overlap {\it locally} over their thermal linewidths. The
approximation is not challenged by thin lines, or by thick lines that
overlap only over the full wind broadening of thousands of km
s$^{-1}$ rather than the tens of km s$^{-1}$ thermal speeds.  The 
amount of local line overlap is
related to the amount of turbulent broadening that is assumed.  In
the presence of very strong redistribution, as assumed in \S \ref{sec:ssr}, 
local line overlap increases the Sobolev optical depths at the expense
of the number of optically thick lines in a complicated way.  Thus
we choose to focus on the possibility of relatively weak turbulence and 
therefore relatively little local line overlap.

\subsection{The Sobolev Approximation}
\label{subsec:sobolev}

The Sobolev approximation asserts that photon interactions with lines
occur primarily due
to Doppler shifts that appear
by virtue of bulk flow velocities, rather than thermal motions.  This
implies that each line interaction is spatially localized.  Due to the
expansion of a stellar wind,
a photon in the comoving frame will
experience continuous Doppler redshifting, loosely analogous to the
Hubble redshift, and will
thus be able to eventually resonate with lines over a much wider
frequency interval than the thermal width.
The fractional Doppler shift
is thus limited only by the scale of the velocity changes $\Delta v$,
set by the terminal speed $v_\infty$.

The infleunce of the thermal speed is to determine the width of
the regions in which a photon may resonate with a given line.
This width is labeled the
Sobolev length $L$ and is given by
\begin{equation}
\label{eq:soblendef}
L = v_{th} \left(\frac{dv}{dr}\right)^{-1} \ ,
\end{equation}
where $v_{th}$ is the thermal speed.
To apply the Sobolev approximation,
we require that the mean free path between resonances with
different lines be much larger than $L$, and this is the sense
in which lines must not overlap.

The primary convenience of this approximation is that it turns the
potentially large number of scatterings within a given resonance
region into a single effective scattering, allowing the radiative
transfer to be modeled as a random walk linking these effective
scatterings.  This type of effective opacity is determined not just by
the density, but also by the velocity gradient, due to the continuous
Doppler shifting referred to above.  The treatment of this
type of opacity was pioneered by CAK, and the resulting description of
the wind dynamics is termed CAK theory.  CAK theory describes a system
in which a typical photon scatters at most once before escaping the
wind.  While adequate for OB stars, the large amount of momentum
deposited in W-R winds requires a typical photon to scatter many times
within the wind.  This ``multiple scattering'' was introduced by
\citet{friend-castor1983}, verified by \citet{springmann1994} using a
Monte Carlo approach, and further refined by \citet{gayley-etal1995}.


\subsection{CAK Theory}
\label{subsec:caksec}

To apply
CAK theory, it is convenient to scale
the radiative acceleration on lines $g_{rad}$
to the radiative acceleration on free electrons $g_e$ by the factor
$M(t)$, called the force multiplier,
\begin{equation}
\label{eq:mglgr}
M(t) = \frac{g_{rad}} {g_e} \ .
\end{equation}
$M(t)$ depends on the optical depths $\tau_{sob}$ across the Sobolev
length $L$ of all the lines in the distribution, which may be
parametrized by $t$, the column depth across $L$ in electron-scattering
optical-depth units:
\begin{equation}
\label{eq:tdef}
t = \rho \kappa_e L \ ,
\end{equation}
where $\kappa_e$ is the free-electron cross-section per gram.
The Sobolev optical depth of line $i$ is given by
\begin{equation}
\label{eq:tausobdef}
\tau_{sob,i} = \kappa_i \rho L \ ,
\end{equation}
where $\kappa_i$ is the cross-section per gram of line $i$ and $\rho$ is
the local density, so
$\tau_{sob,i} \propto t$.

The heart of the CAK approach is to parametrize $M(t)$ by the relation
\begin{equation}
\label{eq:fmdef}
M(t) = k t^{-\alpha},
\end{equation}
where $k$ and $\alpha$ are obtained from the
line-strength distribution.
Specifically, $\alpha$ is a number
between 0 and 1 that relates to the fraction of lines that are
optically thick.
$M(t)$ depends on $t^{-\alpha}$ because the radiative force is saturated
with respect to lines that are already thick, and so increases in $t$
merely dilute the force on such lines over more material.

As an O star loses its hydrogen envelope and evolves into its W-R
phase, its radius shrinks, and this has two immediate consequences for
its wind. First, the smaller radius concentrates the column depth
and  hence increases the optical depth
of a wind of given mass flux.
Second, since W-R stars maintain high luminosity, the reduced radius
implies higher temperatures at the wind base.
The main goal of this work is to understand
how increases in temperature and optical depth affect the star's
capacity for driving mass loss.

\section{Effectively Gray Model}
\label{sec:gray}

An informative first step in such an analysis is to
consider effectively gray opacity, combined with a
non-isotropic diffusion treatment of the radiative transfer.
Indeed, although a
truly gray opacity model requires there be no frequency
dependence at all, in the absence of frequency redistribution
there can be no correlation between the flux $F(\nu)$ and the opacity
$\chi(\nu)$, thus any redistribution-free model is {\em effectively}
gray regardless of the opacity spectrum, neglecting weak
frequency dependences in the diffusivity correction factor $F_{NID}$.
The effectively
gray model then allows CAK-type
theory to apply for appropriately flux-weighted opacities, even in the
limit of multiple scattering in a wind with large total optical depth.


To calculate the momentum
deposition rate for effectively gray opacity, one merely
needs to know the average mean free path {\it between} lines,
rather than within lines.
The effective opacity of the lines
$\chi_{eff}$ is the mean probability of scattering per unit length, and
is given by the expression \citep[e.g.,][]{pinto-eastman2000}
\begin{equation}
\chi_{eff}(\nu) = \frac{dv}{dr}\frac{\nu}{c}\frac{1}{\Delta\nu}
    \sum_i \left(1-e^{-\tau_{sob,\;i}}\right) \ , \label{eq:chieff}
\end{equation}
where $i$ represents a given line and the sum is done over all lines
within an arbitrarily coarse-grained frequency step $\Delta\nu$.   The
term $\left(1-e^{-\tau_{sob,\;i}}\right)$ is
the probability of interacting with line $i$, and
$\Delta\nu$ is selected such that the sum over $i$ equals unity.
When this criteria would require
$\Delta\nu > \nu v_\infty / c$, so it is greater than the Doppler shift
over the whole wind, we truncate $\Delta \nu$ at this value.
To emphasize the role of multiple scattering,
we choose to express the
effective opacity in optical-depth units,
\begin{equation}
\label{eq:effopdepthcalc}
\tau_{eff}(\nu) = v_\infty \frac{dr}{dv} \chi_{eff}(\nu) =
\frac{v_\infty}{c} \frac{\nu}{\Delta\nu} \sum_i \left(1 -
e^{-\tau_{sob,\;i}}\right) \ ,
\end{equation}
such that the opacity parameter $\tau_{eff}(\nu)$
estimates the number of mean free paths
due to line opacity over the full radial extent of the wind.

\subsection{The equations}
\label{subsec:linelist}
In the CAK model the mass-loss rate is set at the critical point,
which essentially occurs at the point where the force efficiency is at a
minimum.  Any material that crosses the critical point will quickly
accelerate, eventually reaching a terminal velocity $v_\infty$.  The
mass-loss rate is assumed to be the maximum value that allows a force
balance at the critical point,
\begin{equation}
\label{eq:forcebaleq1}
v \frac{dv}{dr} = g_{grav} + g_{rad},
\end{equation}
where the force due to gas pressure has been ignored, and $g_{grav}$ is
the acceleration due to gravity:
\begin{equation}
\label{eq:ggravdef}
g_{grav} = -\frac{G \mathcal{M}}{r^2},
\end{equation}
where $G$ is the gravitational constant and $\mathcal{M}$ is the mass of
the star.  The radiative acceleration $g_{rad}$ has two terms, one for
the radiative acceleration from free electrons, and one for the
radiative acceleration from lines:
\begin{equation}
\label{eq:gradeqs}
g_{rad} = g_{e} + g_{L} = \frac{\kappa_e L_*}{4 \pi
r^2 c} \left[1 + M(t) F_{NID} \right],
\end{equation}
Here $F_{NID}$ is a specific correction for nonradial radiation in the
non-isotropic diffusion approximation \citep{gayley-etal1995}, as may be
used in the multiscattering Wolf-Rayet domain, although its value near
0.7 at the critical point also applies for the free-streaming radiation
of optically thin applications.

Substituting eqs. (\ref{eq:ggravdef}) and (\ref{eq:gradeqs}) into
eq. (\ref{eq:forcebaleq1}) gives
\begin{equation}
\label{eq:forcebaleq2}
v\frac{dv}{dr} = -\frac{G\mathcal{M}}{r^2} + \frac{\kappa_e L_*}{4 \pi
r^2 c} \left[1 + M(t) F_{NID} \right].
\end{equation}
It is customary to scale to the effective gravity, which includes both
the true gravity and the
radiative force on free electrons, yielding the dimensionless form
\begin{equation}
\label{eq:critpteq}
1 + y = \frac{\Gamma}{1-\Gamma} M(t) F_{NID}.
\end{equation}
The first term on the left-hand side represents effective gravity, and
the second represents the inertia, scaled as
\begin{equation}
\label{eq:ycdef}
y = \frac{r^2 v dv/dr}{G\mathcal{M}(1-\Gamma)}.
\end{equation}
$M(t) F_{NID}$ is the line force, including both the CAK contribution
$M(t)$ and the multiscattering correction $F_{NID}$
\citep{gayley-etal1995}.  As usual, $\Gamma$ is the Eddington
parameter, given by
the ratio of the radiative force on free electrons to gravity,
\begin{equation}
\label{eq:edparamdef}
\Gamma = \frac{\kappa_e L_*}{4 \pi G \mathcal{M} c}.
\end{equation}
If $\Gamma > 1$, the radiation pressure on the electrons exceeds
gravity and no hydrostatic solution exists.  In a W-R atmosphere
$\Gamma \sim 0.5$, so the atmosphere is static except in regions where
the line opacity augments $\Gamma$, i.e., in the wind.

The relation connecting $t$ to the mass-loss rate $\Mdot$
may be determined by noting that in
spherical symmetry in a steady state,
\begin{equation}
\label{eq:rhodef}
\Mdot = 4 \pi r^2 \rho(r) v(r) \ .
\end{equation}
Substituting eqs. (\ref{eq:soblendef}), (\ref{eq:ycdef}), and
(\ref{eq:rhodef}) into eq. (\ref{eq:tdef}) then yields
\begin{equation}
\label{eq:mdot-t}
\Mdot = \frac{4 \pi G \mathcal{M} (1 - \Gamma) y t}{\kappa_e v_{th}} \ ,
\end{equation}
so maximizing $\Mdot$ amounts to finding the maximum product
$y_c t_c$ at the critical point that allows a force balance to
be achieved.

\subsection{The force multiplier for a real line list}
\label{subsec:realist}

The line list data was taken from the Kurucz list (accessed from the
web, based on \citealt{kurucz1979}), which includes both the
oscillator strengths and their wavelengths. For the analysis in \S
\ref{sec:ssrop}, data from the Opacity Project
\citep{badnell-seaton2003} is used.  Recent studies using more
up-to-date line lists have discovered enhanced opacities due to the
so-called "iron bump" at high temperatures \citep{nugis-lamers2002,
  hillier2003}.  However, this enhancement occurs over a very narrow
temperature range, and there is no mechanism to keep the wind within
this temperature range in the vicinity of the critical point; indeed,
it is the nature of critical points as ``choke points'' 
to avoid regions of extra 
opacity. Instead, the
critical point occurs where the driving efficiency is at a minimum,
so is unlikely to occur within the iron bump.
However, generally elevated opacities at higher temperatures, due
to higher states of Fe ionization (but not a sharp ``bump'' feature),
may give rise to a feedback mechanism, whereby a higher
Sobolev-Rosseland mean opacity increases line blanketing
and leads to a higher temperature, further
enhancing the mass-loss rate, bringing in even higher stages
of Fe and increasing the opacity (Hillier,
private communication).

Line driving depends on the Sobolev optical depths
of these lines, so the density and velocity structure must be
supplied.  Also, the atomic level populations must be determined.   An
LTE code by Ivaylo Mihaylov (private communication) is used to
approximate the level populations, which only requires specification of
the
radiation temperature and the atomic partition
functions. Table \ref{table:ionizstages} shows the highest ionization states used.  
Modifications were made to allow the code to calculate the
resulting effective line optical depth at $10^4$ frequency points
using eq. (\ref{eq:effopdepthcalc}).

\clearpage
\begin{deluxetable}{rrrrrr}
\tablecaption{Highest Ionization Stages of Included Elements
\protect\label{table:ionizstages}}
\tablewidth{0pt}
\tablehead{\colhead{Element} & \colhead{Stage}  & \colhead{Element} &
\colhead{Stage} & \colhead{Element} & \colhead{Stage}}
\startdata
 H &   II & Na & V & Sc &  V\\
He &  III & Mg & V & Ti &  V\\
Li &   IV & Al & V &  V &  V\\
Be &    V & Si & V & Cr &  V\\
 B &    V &  P & V & Mn &  V\\
 C &  VII &  S & V & Fe & IX\\
 N & VIII & Cl & V & Co &  V\\
 O &   IX & Ar & V & Ni &  V\\
 F &    V &  K & V & Cu &  V\\
Ne &   IX & Ca & V & Zn &  V\\
\enddata
\end{deluxetable}
\clearpage

The code is also used to
calculate $M(t)$ given the hydrogen abundance $X$, helium abundance $Y$,
temperature $T$, CAK electron optical depth $t$, and electron number
density $n_e$.  The code was run with $X = 0$ and $Y = 0.98$, to
simulate W-R stars in their helium-dominated ``WNE'' phase. This also
implies that metals comprise the remaining 2\% of the stellar
composition,
a canonical value that is important for line opacity.
The electron number density used in the exitation balance is scaled
proportionally to $t$, such that $n_e = 1\times10^{13} cm^{-3}$ when  $t
= 0.01$, which essentially asserts that our wind model has fixed radius
and velocity and scales that are roughly characteristic of W-R winds.
This results in $n_e$ values that are rather high for O-star winds,
but this is not viewed as a fundamental difficulty as the driving
efficiency is only weakly related to $n_e$.

At a given temperature, the code can be run for several different
values of $t$, which gives $\alpha$ using eq.
(\ref{eq:tdef}).  In CAK theory for an optically think wind irradiated
by a point star, the value of the inertia scaled $y$ at the critical
point is
\begin{equation}
\label{ycritalpha}
y_c = \frac{\alpha}{1 - \alpha} \ ,
\end{equation}
and here this is only slightly modified for nonradial radiation by the
$F_{NID}$ correction factor.
Once $\alpha$ is determined and $y_c$ is found, $\Mdot$ can be
calculated from eq. (\ref{eq:mdot-t}), where $t$ must satisfy
eq. (\ref{eq:critpteq}).

Table \ref{table:mdot-gray} shows the self-consistent mass-loss rates
for gray models at $T = 4\times10^4K$ (a typical O-star temperature) and
$T = 1.3\times10^5K$, a temperature reflecting the smaller radius yet
comparable luminosity of a W-R star. Table \ref{table:assumpt} lists the 
model assumptions. The $4\times10^4K$ model has a mass-loss rate of
about $1.6\times10^{-5}\Msun $yr$^{-1}$, which is large for an O star,
but not as large as many W-R wind projections.
The $1.3\times10^5K$ gray model, however,
yields a mass-loss rate of about $3.0\times10^{-5}\Msun $yr$^{-1}$,
which corresponds to standard clumping-modified
W-R mass-loss rates \citep[such as in][]{hillier-miller1999}.
Thus the gray-opacity analysis reveals
an important piece of the puzzle: the higher-temperature
surfaces of helium stars shifts the stellar luminosity deeper
into the EUV regime where effective line opacity is enhanced.  But the
gray assumption certainly overestimates the line-driving efficiency,
because in reality the flux will tend to be frequency redistributed into
underdense line domains.
To constrain the potential severity of this problem, we now turn to the
opposite limit of extremely efficient frequency redistribution.

\clearpage
\begin{deluxetable}{rrr}
\tablecaption{Model Assumptions
\protect\label{table:assumpt}}
\tablewidth{0pt}
\tablehead{\colhead{$M_*$} & \colhead{$\Gamma$} & \colhead{$\tau_c$}}
\startdata
$12.6\Msun$ & 0.5 & 10\\
\enddata
\end{deluxetable}

\clearpage

\begin{deluxetable}{rrrrrr}
\tablecaption{Parameters and Gray Mass-Loss Rates
\protect\label{table:mdot-gray}}
\tablewidth{0pt}
\tablehead{\colhead{$T$} & \colhead{$\alpha$} & \colhead{$t$} &
\colhead{$M(t)$} & \colhead{$y_c$} & \colhead{$\Mdot (\Msun  yr^{-1})$}}
\startdata
$4.0\times10^4$ & 0.59 & 0.035 & 3.5 & 1.4 & $1.6\times10^{-5}$\\
$1.3\times10^5$ & 0.81 & 0.039 & 7.4 & 4.2 & $3.0\times10^{-5}$\\
\enddata
\end{deluxetable}
\clearpage

\section{Results for the SSR Model}
\label{sec:ssr}

The limit of efficient frequency redistribution over a highly diffusive
radiation field in a supersonically expanding wind allows the
application of the statistical Sobolev-Rosseland (SSR) opacity
approximation \citep{onifer-gayley2003}. This approximation asserts that
completely redistributed source functions give rise to a diffusive flux
that is inverse to the local frequency-dependent effective opacity, as
holds for the Rosseland mean in static photospheres, but the effective
opacity is controlled by the Sobolev approximation and is governed by
eq. (\ref{eq:effopdepthcalc}).
Since the radiative acceleration is governed by the frequency-integrated
product of flux times opacity, the inverse correlation between them has
important implications, and CAK theory must be augmented by a
consideration of the line frequency distribution, not just the
line-strength distribution.

Since we wish to focus on the frequency dependence of the flux that is
caused by the frequency dependence of the opacity, it is convenient to
transform to a new frequency-like variable, such that the flux
arising from truly gray opacity would be {\it flat} when
expressed in terms of this variable.
This may be accomplished by introducing
the flux ``filling factor'' $f$ given by
\begin{equation}
f \ =  \ \frac{\int_{\nu_{min}}^{\nu} B(\nu')
d\nu'}{\int_{\nu_{min}}^{\nu_{max}} B(\nu) d\nu} \ ,
\label{eq:fdef}
\end{equation}
where $B(\nu)$ is the envelope function expressing the gross
frequency dependence of the stellar flux for gray opacity (approximated
here by the Planck function for simplicity).
Thus the flux
density, per interval of $f$ instead of $\nu$, is constant
everywhere over $f$-space in the absence of non-gray opacity
modifications, and hence provides a useful space to characterize such
modifications.
Regions in frequency space that have a low
incoming flux density map into narrow regions in $f$ space, and regions
of frequency space that have a large incoming flux density map into wide
regions in $f$ space (see Figure
\ref{fig:4panel1}a-d).
The large gap seen on the
left-hand side of Figure \ref{fig:4panel1}a occurs at a region
in frequency space with a low incoming flux density, as seen in Figure
\ref{fig:4panel1}b.  Thus it is mapped into a small sliver of
$f$ space, as seen in Figure
\ref{fig:4panel1}c.  The opacity spike seen near log$(\nu) =
15.9$ occurs at the peak of the Planck curve, and therefore is mapped
into a wide region of $f$ space.  Figure \ref{fig:4panel1}d
exhibits the expected flat
profile when the frequency dependence of the flux follows $B(\nu)$ from
Figure \ref{fig:4panel1}b.

The primary advantage of working in $f$-space is the convenience of
calculating the radiative acceleration from lines, $g_L$, which
involves flux weighting
the opacity function $\tau(f)$ (here expressed in optical depth units as
per eq. [\ref{eq:effopdepthcalc}]).
The flux-weighted result is
\begin{equation}
\label{force-fspace}
g_L \ \propto \ \int_0^1 F(f) \tau_L(f) df \ ,
\end{equation}
where $F(f)$ describes the frequency dependence of the flux function,
and use of $f$-space allows the structure of $F(f)$ to be induced
by $\tau_L(f)$ independently from the frequency dependence of the
sources. In particular, when redistribution is neglected, $F(f)$ is flat
and the radiative acceleration is determined by the simple integral of
$\tau_L$, whereas in the opposite SSR limit of extreme
redistribution, $F(f)$ is inversely proportional to $\tau_L(f)$
and the entire integrand of eq. (\ref{force-fspace}) becomes flat.

The form of eq. (\ref{force-fspace}) makes it evident that $g_L$ depends
on appropriate mean opacities.
If the goal is to contrast the SSR approximation with the gray result,
the relevant effective line-opacity means, in optical-depth units, are
the gray average $\tau_g$, the SSR flux-weighted mean $\tau_{SSR}$
(including the impact of lines plus continuum),
and the pure continuum mean $\tau_c$ (assumed here to be manifestly
gray). Note that $\tau_g$
and $\tau_{SSR}$ can be determined from the line list and have
the functional form:
\begin{eqnarray}
\tau_g & = & \int_0^1 \tau_L(f) df, \label{eq:taugdef} \\
\tau_{SSR} & = & \left[\int_0^1 \left(\tau_L(f) + \tau_c\right)^{-1}
df\right]^{-1} \ , \label{eq:taussrdef}
\end{eqnarray}
where the latter expresses the abovementioned inverse scaling of flux
and opacity, in a manner entirely analogous to the static Rosseland mean
but typically generating far larger line opacities as a result of the
supersonic flow.

A key quantity that depends only on these mean opacities is the force
efficiency $\mathcal{E}$, defined as
the ratio of the actual line force to the line force
that would result for a flat $F(f)$ that did not respond to the opacity,
i.e., for an effectively gray force.
When the line acceleration is treated in the SSR approximation, this
yields \begin{equation}
\label{eq:forceffgl}
\mathcal{E} \ = \ \frac{g_{SSR}}{g_{gray}} \ ,
\end{equation}
and removing the continuum opacity to yield line accelerations in units
of mean optical depths gives
\begin{eqnarray}
g_{gray} & \propto & \tau_{gray} - \tau_c \ = \ \tau_g \label{eq:taug}\\
g_{SSR} & \propto & \tau_{SSR} - \tau_c \ , \label{eq:glssr}
\end{eqnarray}
such that
\begin{equation}
\label{eq:forceffdef}
\mathcal{E} \ = \  \frac{\tau_{SSR} - \tau_c}{\tau_g} \ ,
\end{equation}
where $\tau_g$, $\tau_{SSR}$, and $\tau_c$ are as defined above.


An additional simplification may now be added to the analysis.
Once the degree of (anti)correlation between $\tau_L(f)$ and $F(f)$ is
specified, it is no longer necessary for the purposes of eq.
(\ref{force-fspace}) that the proper {\it sequence} in $f$-space be
maintained, and
the frequency-dependent opacities and fluxes may be reordered
arbitrarily into some other $\fprime$ space so long as the mapping from
$df$ to $d\fprime$ has unit Jacobian. We choose to sort the opacity
distribution in decreasing order of $\tau_L$ (see Figure
\ref{fig:4panel2}), which permits a {\it monotonic} $\tau_L(\fprime)$
distribution over the resorted $\fprime$ space.
The new distribution over $\fprime$ generates a new $F(\fprime)$, but
$g_L$ is of course preserved by this simple change of integration
variable.
The advantage of monotonic $\tau_L(\fprime)$ and $F(\fprime)$ is that
they may be approximated by analytic curves, and the properties of those
analytic fits offer insights that are not available from a direct
numerical evaluation of $g_L$.

Figure \ref{fig:4panel2} shows the opacity when sorted in decreasing
order over $\fprime$-space. We approximate the resulting smooth
opacity curve by an exponential of the form
\begin{equation}
\label{eq:expfiteqn}
\tau(\fprime) \ = \ a e^{-b\fprime} \ ,
\end{equation}
where $b$ parameterizes the level of nongrayness.
If $b = 0$ the opacity is the same at all frequencies, and the lack of
any gaps implies that this gray result is the most efficient for driving
the wind. As $b$ increases, the importance of gaps increases,
and the wind driving efficiency and mass-loss rate drop.
For large $b$, the exponential falls so rapidly that it generates a
frequency domain that is nearly line free,
the ramifications of which are considered in \citet{onifer-gayley2003}.
The operational values of the opacity scale parameter $a$ and
nongrayness parameter $b$ are chosen to exactly recover both the gray
opacity and the SSR force efficiency $\mathcal{E}$ from numerical
integrations.

In the SSR approximation, the diffusive correction $F_{NID}$ is
further altered by the non-gray correction $\mathcal{E}$, such that eq.
(\ref{eq:critpteq}) is replaced by
\begin{equation}
\label{eq:ssrcritpnt}
1 \ + \ y_c \ = \ \frac{\Gamma}{1-\Gamma} M(t) F_{NID} \;\mathcal{E}.
\end{equation}
$\mathcal{E}$ is then calculated from eq. (\ref{eq:forceffdef}), where
$\tau_g$ and $\tau_{SSR}$ are calculated from the line list.
The value of $y$ at the critical
point, $y_c$, is calculated by setting to zero the derivative with
respect to $y$ of eq. (\ref{eq:ssrcritpnt}) at $y = y_c$.
Since $M(t)$ is
the ratio of the gray line force to the free electron force,
\begin{equation}
\label{eq:fmtgte}
M(t) \ = \ \frac{\tau_g}{\tau_e} \ ,
\end{equation}
where we assume $\tau_e = \tau_c$. We apply the CAK ansatz to obtain
\begin{equation}
\label{eq:mpropy}
M(t) \ \propto \ y^{\alpha} \ \propto \ t^{-\alpha} \ .
\end{equation}
Substituting eq. (\ref{eq:expfiteqn}) into eqs.
(\ref{eq:taugdef}) and (\ref{eq:taussrdef}) then gives
\begin{equation}
\tau_g \  =  \ \frac{a}{b} \left( 1 - e^{-b} \right),
    \label{eq:taugint}
\end{equation}
and
\begin{equation}
\tau_{SSR} \ = \ \frac{b \tau_c}{ln \left(\frac{b M + e^b - 1}{b M -
    e^{-b} + 1} \right)} \ , \label{eq:taussrint}
\end{equation}
where $M = M(t)$ and eqs. (\ref{eq:fmtgte}) and
(\ref{eq:taugint}) have been used to eliminate $a$.
Equations (\ref{eq:taugint}) and
(\ref{eq:taussrint}) may then be substituted into eq.
(\ref{eq:forceffdef}), yielding
\begin{equation}
\label{eq:febm}
\mathcal{E} \ = \ M^{-1} \left[ \frac{b}{ln \left(\frac{bM + e^b - 1}{bM
- e^{-b} + 1}\right)} - 1 \right] \ .
\end{equation}
To find the values of $y_c$ and $M_c$ at the critical point,
we substitute eqs. (\ref{eq:febm}) and (\ref{eq:mpropy})
into eq. (\ref{eq:ssrcritpnt}) and set the derivative at $y_c$ to
zero, yielding
\begin{equation}
\label{eq:critptcond}
\frac{\Gamma' F_{NID} b^2 M_c \alpha y_c^{-1}
 \left[ \left(b M_c - e^{-b} +
1 \right)^{-1} - \left(b M_c + e^b - 1\right)^{-1}\right]}
{\left[ln \left(\frac{b M_c + e^b - 1}{b M_c - e^{-b}
+ 1} \right) \right]^{2}} \ = \ 1 \ ,
\end{equation}
where $\Gamma' \ = \ \Gamma / (1 - \Gamma)$.
A second constraint on $M_c$ is obtained by solving
eq. (\ref{eq:ssrcritpnt}), after using eqs.
(\ref{eq:febm}) and (\ref{eq:mpropy}), so
\begin{equation}
\label{eq:hdef}
M_c \  = \ \frac{e^{\left(\frac{b}{1 + (1 + y_c) / \Gamma'
F_{NID}}\right)} \left(1 - e^{-b}\right) + 1 - e^b}{\left[1 -
e^{\left(\frac{b}{1 + (1 + y_c) / \Gamma'
F_{NID}}\right)}\right] b}.
\end{equation}
Equations (\ref{eq:critptcond}) and (\ref{eq:hdef}) can be combined and
$y_c$ and $M_c$ found numerically.

The line list is analyzed with specified $t$, $\tau_c$, $T$, and
$n_e$, which yield $\tau_g$, $\tau_{SSR}$, $\mathcal{E}$, and $b$. The
CAK $\alpha$ is determined the same way as in the gray case, by varying
$t$ and seeing its effect on $M_c$, where $\alpha = -d$ln $M / d$ln $t$.

The critical point is
found numerically assuming
$F_{NID} = 0.7$, and eq. (\ref{eq:ssrcritpnt}) is then checked for
consistency.
If $1 + y_c \ne \Gamma' F_{NID} M(t) \mathcal{E}$, then
the procedure is repeated with an updated $t$ until eq.
(\ref{eq:ssrcritpnt}) is self-consistent.
Equation (\ref{eq:mdot-t})
then gives the mass-loss rate.

\subsection{Results and Discussion}
\label{sec:ssrmdotsubsec}

\begin{figure}
\plotone{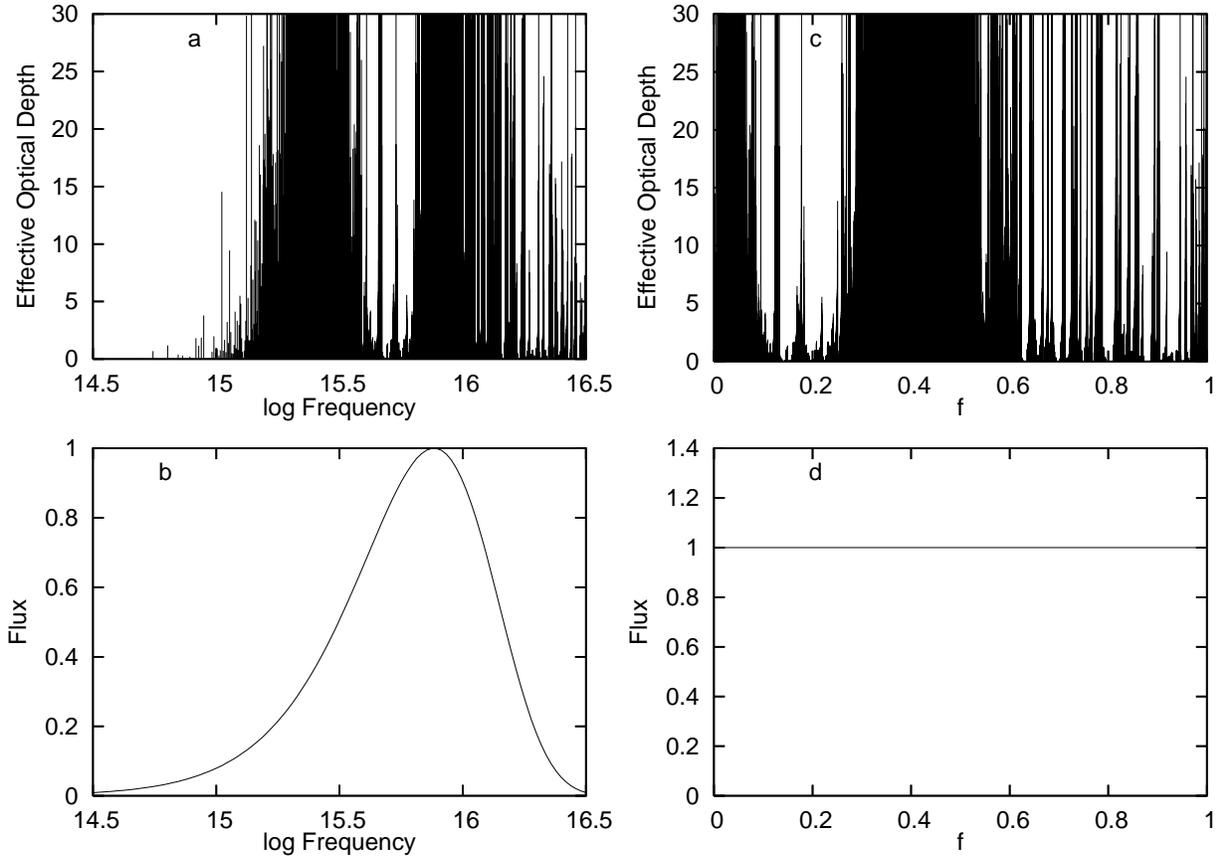}
\caption{The conversion of frequency space to flux filling factor
space.  Panels (a) and (b) show the opacity and the incoming flux,
respectively, as a function of the frequency.  Panels (c) and (d) show
the same opacity and incoming flux in flux-filling-factor space. The
temperature is 130,000K. \protect\label{fig:4panel1}}
\end{figure}

\begin{figure}
\plotone{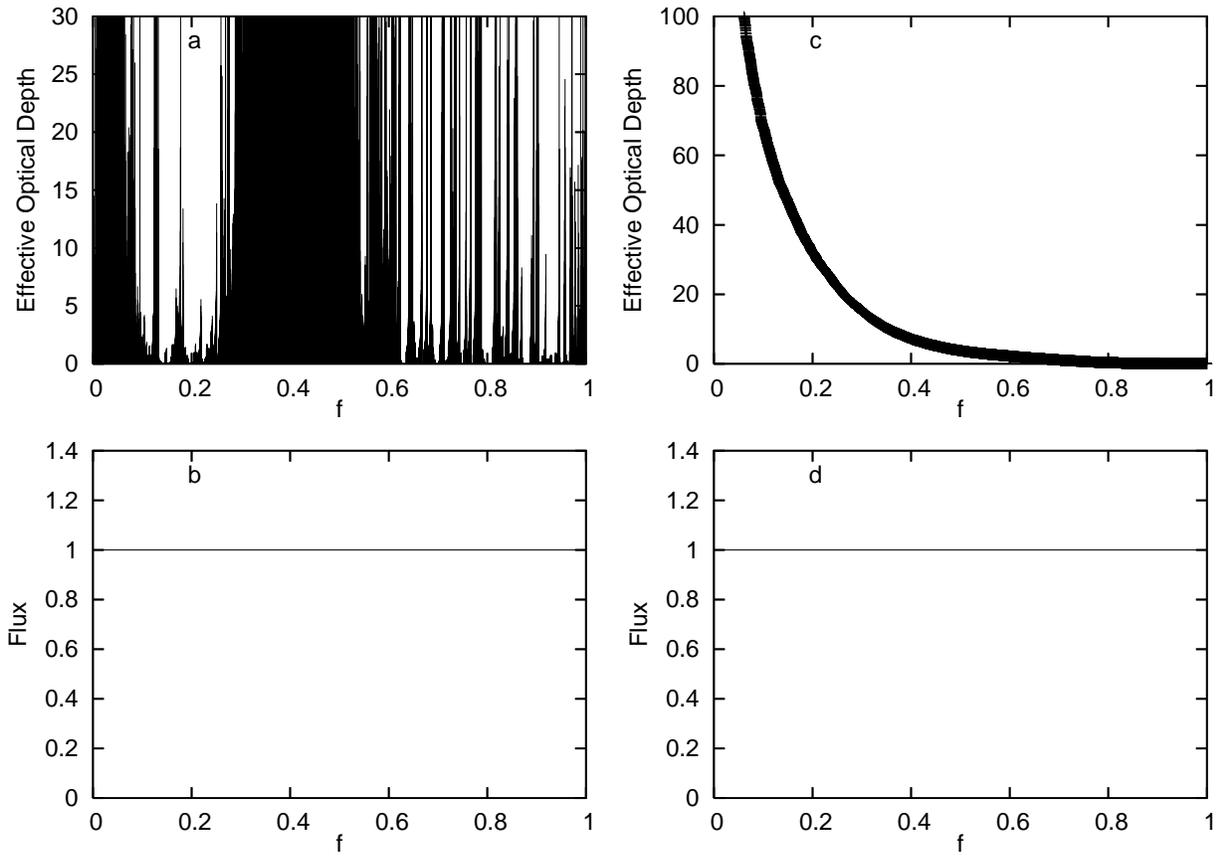}
\caption{The effective optical depth (a)
is sorted in decreasing order (b).  This produces the same force,
since the incoming flux (c and d) is flat in f-space.
\protect\label{fig:4panel2}} \end{figure}

The SSR model is most appropriate for a wind that is highly
redistributing and optically thick, such that
photons are quickly shunted into spectral gaps where long
mean-free-paths enables them to carry much of the stellar flux, and
force
efficiency drops.
Therefore, redistribution significantly reduces the
mass-loss rate relative to gray scattering.
Figures
\ref{fig:4panel1} and \ref{fig:4panel2} show the effective line
optical depth spectrum of an LTE wind with $T = 1.3\times10^5K$ at the
critical point.
The last row in Table \ref{table:mdot-ssr} shows
the parameters that result from such a model, where we have chosen a
characteristic value of the Eddington parameter $\Gamma = 0.5$ (and 
the other model parameters listed in Table \ref{table:assumpt}).
The force efficiency
drops to about 23\% of the gray force efficiency, which translates into
a mass-loss rate of $4.3\times10^{-6}$.
This is
significantly below observed W-R mass-loss rates, even with
clumping corrections \citep[e.g.,][]{nugis-etal1998,
hillier-miller1999}. Thus we conclude that if the SSR limit of extreme
frequency
redistribution applies locally in W-R winds, then
line-driving theory cannot explain their
mass-loss rates using the Kurucz line list.
To increase the mass-loss rate via line driving theory, it would be
necessary to either fill the gaps by including additional lines, or to
include the finite time required to redistribute
photons into the opacity gaps by relaxing the CRD approximation, thus
allowing photons to scatter multiple times before a redistribution
occurs.


\clearpage

\clearpage

\clearpage

\begin{deluxetable}{rrrrrrrrrr}
\tablecaption{Gray and SSR Mass-Loss Rates From the Kurucz List
\protect\label{table:mdot-ssr}} \tablewidth{0pt}
\tablehead{\colhead{$T(K)$} & \colhead{Type} & \colhead{$\alpha$}
& \colhead{$t$} & \colhead{$M(t)$} & \colhead{$\mathcal{E}$}
& \colhead{$a$} & \colhead{$b$} & \colhead{$y_c$} & \colhead{$\Mdot (\Msun yr^{-1})$}}
\startdata
$1.3\times10^5$ & Gray & 0.81 & 0.039 & 7.4 & 1.0 &
\multicolumn{1}{c}{--} & \multicolumn{1}{c}{--} & 4.3 & $3.0\times10^{-5}$\\
$1.3\times10^5$ & SSR & 0.79 & 0.016 & 15 & 0.23 & 14 & 5.2 & 1.5 &
$4.3\times10^{-6}$\\
\enddata
\end{deluxetable}

\clearpage

\section{The Opacity Project Data}
\label{sec:ssrop}

\begin{figure}
\plotone{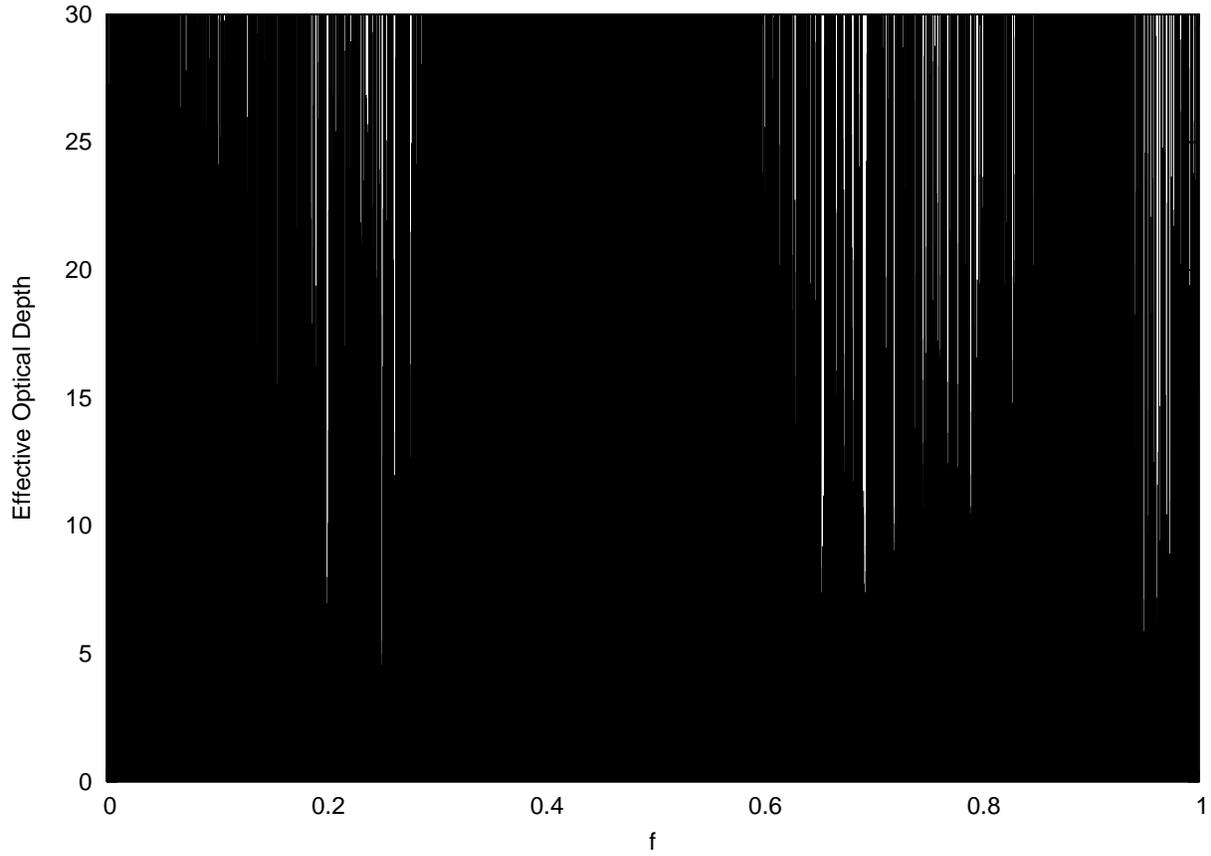}
\caption{The effective optical depth as a function of the flux filling
  factor f due to the Opacity Project line list.  Compare to figure
  \ref{fig:4panel1}c \protect\label{fig:op1}}
\end{figure}

The most obvious way to fill opacity gaps is to find new opacity.  To
that end, we have replaced the Kurucz oscillator strengths with
oscillator strengths from the more complete Opacity Project (OP)
\citep{badnell-seaton2003}.  Kurucz oscillator strengths were used for
P, Cl, K, Sc, Ti, V, Cr, Mn, Co, Ni, Cu, and Zn, as these elements are
not available through the Opacity Project.  In addition, we have
raised the highest stage of iron to XIII.  As before, the temperature
is $1.3\times10^4K$.  Figure \ref{fig:op1} shows the effective optical
depth as a function of f (compare to figure \ref{fig:4panel1}c).  The
OP list contains about twice as many lines within the wavelength range
and ionization states of table \ref{table:ionizstages} as the Kurucz
list.  As shown in Table \ref{table:mdot-ssr-op}, the SSR mass-loss rate is $\Mdot = 7.0\times10^{-6}$, about
twice as large as the Kurucz value, but still insufficient if line
driving is to describe all but the weakest W-R winds.  To acheive
higher mass-loss rates within the CAK regime, another method is needed
to introduce additional lines.

\clearpage

\clearpage

\begin{deluxetable}{rrrrrrrrrr}
\tablecaption{Gray and SSR Mass-Loss Rates From the OP List
\protect\label{table:mdot-ssr-op}} \tablewidth{0pt}
\tablehead{\colhead{$T(K)$} & \colhead{Type} & \colhead{$\alpha$}
& \colhead{$t$} & \colhead{$M(t)$} & \colhead{$\mathcal{E}$}
& \colhead{$a$} & \colhead{$b$} & \colhead{$y_c$} & \colhead{$\Mdot (\Msun yr^{-1})$}}
\startdata
$4.0\times10^4$ & Gray & 0.55 & 0.029 & 3.16 & 1.0 &
\multicolumn{1}{c}{--} & \multicolumn{1}{c}{--} & 1.2 & $1.2\times10^{-5}$\\
$1.3\times10^5$ & Gray & 0.73 & 0.13 & 5.24 & 1.0 &
\multicolumn{1}{c}{--} & \multicolumn{1}{c}{--} & 2.7 & $6.2\times10^{-5}$\\
$1.3\times10^5$ & SSR & 0.72 & 0.035 & 13.5 & 0.22 & 332 & 5.3 & 1.1 &
$7.0\times10^{-6}$\\
\enddata
\end{deluxetable}

\section{Ionization Stratification}
\label{sec:ionstrat}

One way to fill opacity gaps {\it nonlocally}, thereby
increasing the force efficiency, was
suggested by \citet[][hereafter LA93]{lucy-abbott1993} and
involves the appearance of a
large number of additional lines via ionization stratification.
Such an ionization gradient has been observed to be fairly ubiquitous in
W-R stars \citep{herald-etal2000}, and in the inner wind appears due to the
significant temperature drop over the span of the optically thick wind
envelope.  When such a gradient exists at the spatial scale over which
photons diffuse prior to being redistributed in frequency, local gaps
left by
one ionization state  of a given element may be filled
by lines of another state located nearby.
This in effect creates a globally gray line list and mitigates any local
nongrayness, but is only effective over the length scale of frequency
thermalization. Thus the extreme redistribution limit would still reduce
to the local SSR approximation, but finite thermalization lengths would
yield results that are intermediate to the widely differing gray and SSR
mass-loss rates derived above.

\subsection{Two-Domain Model With Ionization Stratification}
\label{2domionstratsubsec}

To obtain conceptual insight into the effect of ionization
stratification on the overall mass-loss rate, we revisit the model of
\citet{onifer-gayley2003}, where the line list is replaced by a simple
model containing two frequency domains.
One domain contains effective
line opacity $\tau_{L1}$, while the other contains effective line
opacity $\tau_{L2}$, both in optical depth units as usual.
There is also a continuum opacity
$\tau_c$ that spans both frequency domains.
This model allows us to identify important characteristics
of the system separately from the details of the line list, and is
simple enough to permit analytic approximation even in the presence of
spatial variations.

\begin{figure}
\plotone{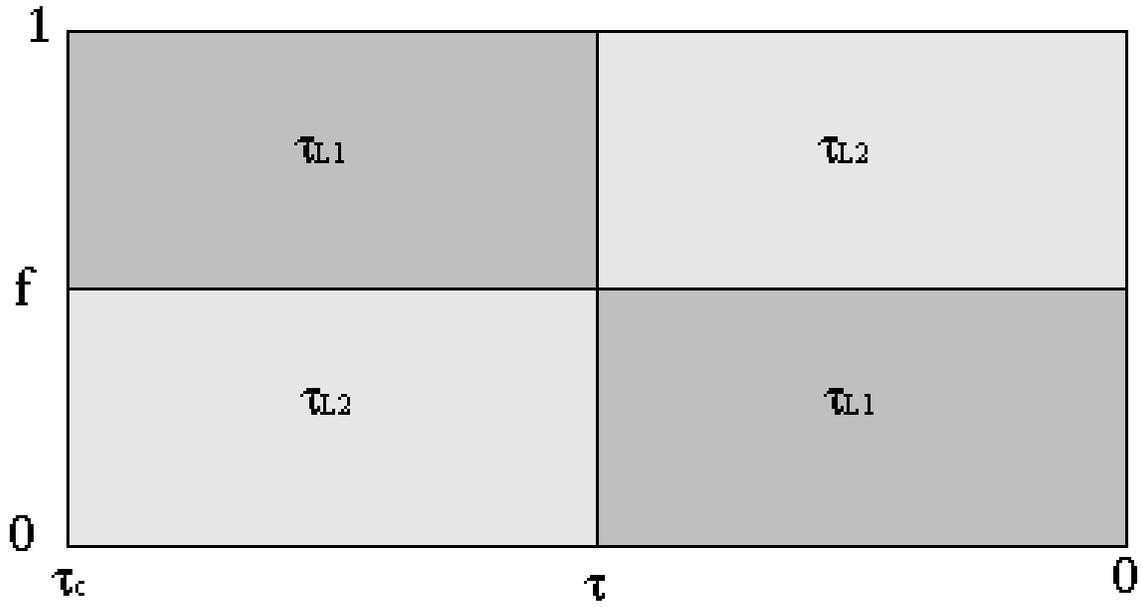}
\caption{The ionization stratification
model.  The y-axis represents f-space and the x-axis is measured in
units of the continuum optical depth, from $\tau = \tau_c$ at the left
end to $\tau = 0$ at the right end.  The radiation from the star
enters the wind from the left. \protect\label{fig:toymodel}}
\end{figure}

To account in a simple conceptual way
for ionization stratification and its potential for filling
the gaps globally, an additional complication is introduced to the
plane-parallel model atmosphere that is used to signify the wind, as
illustrated in Figure \ref{fig:toymodel}. In addition to dividing the
frequency domain into two equal parts (i.e., with $f = 1/2$) and
supplying them with line optical depths $\tau_{L1}$ and $\tau_{L2}$, the
atmosphere is also divided equally into two physical-space regions,
between which the opacities in each frequency domain are {\it
interchanged}. The total continuum optical depth $\tau_c$ pervades all
domains and all regions, such that $\tau_c/2$ is the midpoint of the
atmosphere where the opacity interchange is imposed, and the total optical depth
within a given ionization zone and frequency domain $i$ is $\tau_i = \tau_{Li} + \tau_c / 2$.
This yields a kind of toy model of a wind that is globally gray (as the
total optical depth in both frequency domains is the same), but can be
very nongray locally in each region.
For conceptual purposes, the radiative transfer is treated in
the two-stream approximation,
so that the mean intensity and flux are given in terms of intensity
streams in the inward and outward directions:
\begin{eqnarray}
J_\nu & = & (I_{\nu+} + I_{\nu-}) / 2, \label{eq:jdef}\\
F_\nu & = & I_{\nu+} - I_{\nu-} \ . \label{eq:fluxdef}
\end{eqnarray}
The depth variable is chosen to be the continuum optical depth,
\begin{equation}
\label{eq:taudef}
d\tau \ \equiv \ d\tau_c \ = \ \frac{\tau_c}{\tau_i} d\tau_i \ ,
\end{equation}
where $i$ is 1 or 2 to represent the two frequency domains.


\clearpage
\clearpage

To complete the radiative transfer solution for such an atmosphere, it
is necessary to specify the radiation sources.
Here we assume pure scattering in radiative equilibrium, but as
mentioned above, a key issue is the degree of frequency redistribution
per scattering. 
we assume for simplicity that the continuum opacity scatters 
coherently.
\citet{pinto-eastman2000} showed that large 
amounts of line redistribution have a similar 
effect on the overall frequency dependent envelope of the radiative
flux as does continuum redistribution. We allow the line opacity in one domain
to redistribute photons into line opacity in the other domain, using a
simple probabilistic approach. 

To seek an extreme redistribution limit in such a model is to assert that all
line scattering is completely redistributing, so that
the probability that any scattering event within a given ionization stratum
will be redistributing is
\begin{equation}
\Lambda_{1} \ = \ \frac{\tau_{L1}}{\tau_{L1} + \tau_c / 2}
    \label{eq:lambda1def}
\end{equation}
for frequency domain 1 and
\begin{equation}
\Lambda_{2} \ = \ \frac{\tau_{L2}}{\tau_{L2} + \tau_c / 2}
    \label{eq:lambda2def}
\end{equation}
for frequency domain 2, where $\tau_c / 2$ is the continuum optical depth 
within a single ionization zone. Thus if the line opacity $\tau_{Li}$ goes to 0, 
no redistribution 
may occur in that domain, whereas
if $\tau_{Li}$ gets very large, redistribution becomes certain.

However, not all redistribution will result in changing frequency
domain, this must be accounted for with a separate probability that
obeys the requirement of reciprocity and is independent of the original
state of the photon. Thus the probability of redistributing into a
particular frequency domain, given that a redistribution has occurred,
is
\begin{equation}
w_{1} \ = \ \frac{\tau_{L1}}{\tau_{L1} + \tau_{L2}} \label{eq:w1def}
\end{equation}
for redistribution into domain 1, and
\begin{equation}
w_{2} = \frac{\tau_{L2}}{\tau_{L1} + \tau_{L2}} \label{eq:w2def}
\end{equation}
for domain 2. Therefore, the joint probability that a photon will scatter in frequency
domain $i$, redistribute, and result in frequency domain $j$ is $\tau_i \Lambda_i w_j$.
Note that if $\tau_{Li}$ goes to 0, $w_i$ also goes to
0, and no redistribution can occur into that domain.
What is less obvious is that if $\tau_{Lj}$ goes to 0, then $w_i$ goes
to 1, but $\Lambda_j = 0$, so the joint probability $\Lambda_j w_i$ of
redistributing from frequency domain $j$ into frequency domain $i$,
given that a scattering has occured in domain $j$, is still 0.

The fact that eqs. (\ref{eq:lambda1def})-(\ref{eq:w2def}) obey
reciprocity may be seen from
\begin{equation}
\label{eq:jointprobratio}
\frac{p_{12}} {p_{21}}  \ = \  \frac{\tau_1 \Lambda_1
w_2} {\tau_2 \Lambda_2 w_1} \ = \ 1 \ ,
\end{equation}
where $p_{ij}$ is the complete probability of scattering in frequency
domain $i$ and redistributing
into domain $j$.
This condition is all that is required for eqs.
(\ref{eq:lambda1def})-(\ref{eq:w2def}) to yield the
SSR limit if all optical depths are sufficiently large and all gradients
are sufficiently gradual.
However, it is exactly the impact of more rapid gradients that is being
explored by this simple model.

The radiative transport equations that must be solved are
\begin{eqnarray}
\frac{dF_{\ell1}(\tau)}{d\tau} & = & 2 \frac{\tau_1}{\tau_c} \Lambda_1
w_2
    \left( J_{\ell1}(\tau) - J_{\ell2}(\tau) \right), \label{eq:rteq1}\\
\frac{dF_{\ell2}(\tau)}{d\tau} & = & 2 \frac{\tau_2}{\tau_c} \Lambda_2
w_1
    \left( J_{\ell2}(\tau) - J_{\ell1}(\tau) \right), \label{eq:rte12}\\
\frac{dF_{r1}(\tau)}{d\tau} & = & 2 \frac{\tau_1}{\tau_c} \Lambda_1 w_2
\left(
    J_{r1}(\tau) - J_{r2}(\tau) \right), \label{eq:rteq3}\\
\frac{dF_{r2}(\tau)}{d\tau} & = & 2 \frac{\tau_2}{\tau_c} \Lambda_2 w_1
\left(
    J_{r2}(\tau) - J_{r1}(\tau) \right), \label{eq:rteq4}\\
\frac{dJ_{\ell1}(\tau)}{d\tau} & = & \frac{1}{2} \frac{\tau_1}{\tau_c}
    F_{\ell1}(\tau), \label{eq:rteq5}\\
\frac{dJ_{\ell2}(\tau)}{d\tau} & = & \frac{1}{2} \frac{\tau_2}{\tau_c}
    F_{\ell2}(\tau), \label{eq:rteq6}\\
\frac{dJ_{r1}(\tau)}{d\tau} & = & \frac{1}{2} \frac{\tau_1}{\tau_c}
    F_{r1}(\tau), \label{eq:rteq7}\\
\frac{dJ_{r2}(\tau)}{d\tau} & = & \frac{1}{2} \frac{\tau_2}{\tau_c}
    F_{r2}(\tau), \label{eq:rteq8}
\end{eqnarray}
where $\ell$ refers to the left side of the configuration space in
Figure \ref{fig:toymodel} and $r$ refers to the right, and 1 denotes the
frequency domain that initially contains $\tau_{L1}$ while 2 denotes the
frequency domain that initially contains $\tau_{L2}$.
These equations may be recast by the
following substitutions:
\begin{eqnarray}
x_\ell(\tau) & \equiv & J_{\ell2}(\tau) - J_{\ell1}(\tau) \ ,
    \label{eq:xldef}\\
x_r(\tau) & \equiv & J_{r2}(\tau) - J_{r1}(\tau) \ , \label{eq:xrdef}\\
y_\ell(\tau) & \equiv & \frac{\tau_c}{\tau_1} J_{\ell1}(\tau) +
    \frac{\tau_c}{\tau_2} J_{\ell2}(\tau) \ , \label{eq:yldef}\\
y_r(\tau) & \equiv & \frac{\tau_c}{\tau_1} J_{r1}(\tau) +
    \frac{\tau_c}{\tau_2} J_{r2}(\tau) \ , \label{eq:yrdef}
\end{eqnarray}
and then eqs. (\ref{eq:rteq1})-(\ref{eq:rteq8}) become
\begin{eqnarray}
\frac{d^2x_{\ell}(\tau)}{d\tau^2} & = & a x_{\ell}(\tau) \ ,
    \label{eq:rtxleq}\\
\frac{d^2x_r(\tau)}{d\tau^2} & = & a x_r(\tau) \ , \label{eq:rtxreq}\\
\frac{dy_{\ell}(\tau)}{d\tau} & = & F_{\ell,tot} \ , \label{eq:rtyleq}\\
\frac{dy_r(\tau)}{d\tau} & = & F_{r, tot} \ , \label{eq:rtyreq}
\end{eqnarray}
where
\begin{equation}
\label{eq:adef}
a \ = \  \left( \frac{\tau_2}{\tau_c} \right)^2 \Lambda_2 w_1 + \left(
    \frac{\tau_1}{\tau_c} \right)^2 \Lambda_1 w_2 \ .
\end{equation}
The solutions are
\begin{eqnarray}
x_{\ell}(\tau) & = & C_{\ell+} e^{\sqrt{a}\tau} + C_{\ell-}
    e^{-\sqrt{a}\tau} \ , \label{eq:xlsoln}\\
x_r(\tau) & = & C_{r+} e^{\sqrt{a}\tau} + C_{r-} e^{-\sqrt{a}\tau} \ ,
    \label{eq:xrsoln}\\
y_{\ell}(\tau) & = & B_{\ell} + F_{\ell,tot} \tau \ ,
\label{eq:ylsoln}\\ y_r(\tau) & = & B_r + F_{r, tot} \tau \ .
\label{eq:yrsoln}
\end{eqnarray}

The eight constants $C_{\ell/r,1/2}, B_{\ell/r}, F_{\ell/r,tot}$ are
found by using the following eight boundary conditions: ({\it i}) No
photons enter the right side of the wind,
\begin{equation}
\label{eq:bc12}
I_{r-i}(0) = J_{ri}(0) - \frac{1}{2}F_{ri}(0),
\end{equation}
where $i$ is 1 or 2, so this represents two boundary conditions. ({\it
ii}) The fluxes and the mean intensities are
continuous at the midpoint of the wind,
\begin{eqnarray}
F_{\ell i}(\tau_c/2) & = & F_{ri}(\tau_c/2), \label{eq:bc34}\\
J_{\ell i}(\tau_c/2) & = & J_{ri}(\tau_c/2), \label{eq:bc56}
\end{eqnarray}
which gives four more boundary conditions.
({\it iii}) The stellar surface at the left of the atmosphere model is
assumed to be highly redistributing, so that the incoming
intensities on the left are unity,
\begin{equation}
\label{eq:bc78}
I_{\ell+i}(\tau_c) = J_{\ell i}(\tau_c) + \frac{1}{2}F_{\ell i}(\tau_c)
     \ =  \ 1 \ ,
\end{equation}
which gives the final two.


\clearpage
\begin{figure}
\plotone{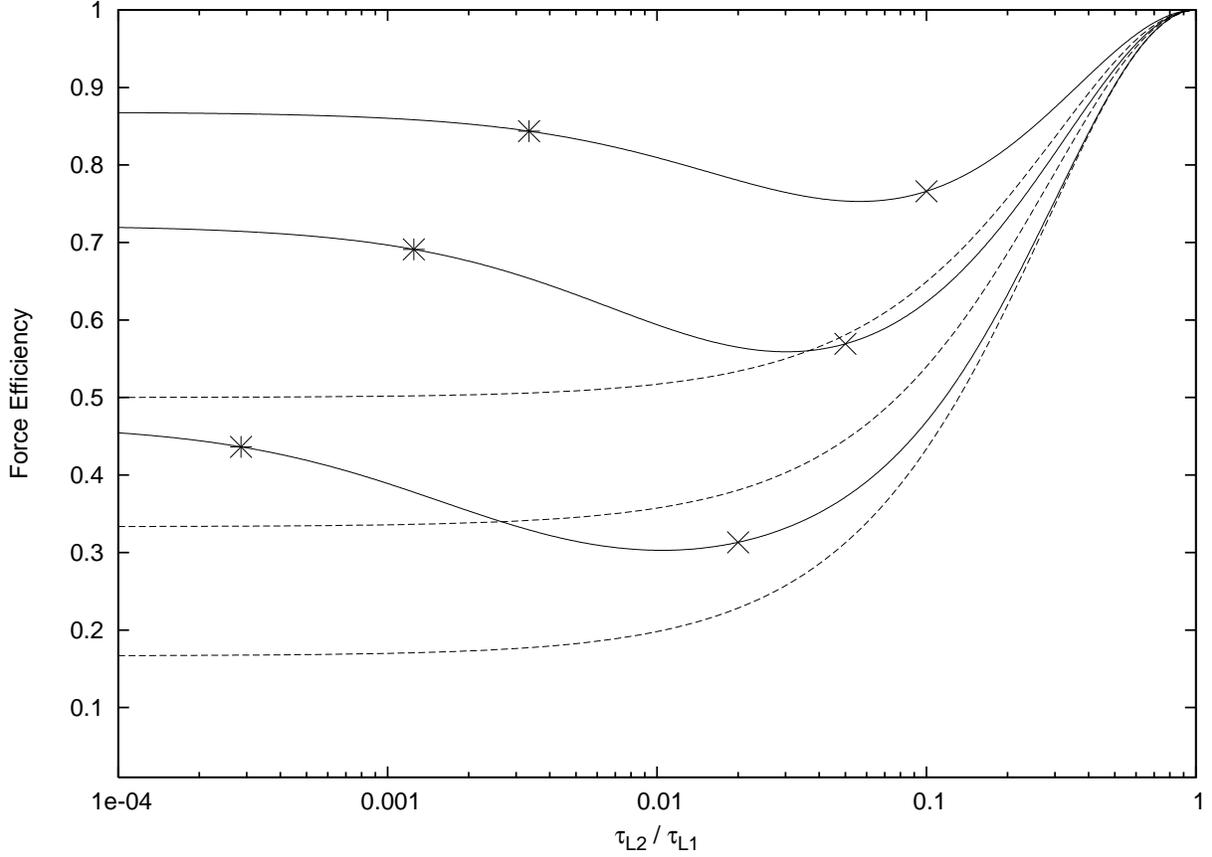}
\caption{
The force efficiency as a function of the opacity ratio at
$\tau = 9$, i.e., one continuum mean-free path from the left edge.
The solid lines are the toy model calculations and the dashed lines
are the SSR results.  The topmost solid and dashed curves are for
$\tau_{L1} = 10$, the middle are for $\tau_{L1} = 20$, and 
the bottom are for $\tau_{L1} = 50$.
The asterisks appear where the atmosphere becomes effectively thick
to redistribution, i.e. $\tau_c = 1/\sqrt{a}$, and the crosses
appear where all frequency domains become optically thick,
i.e., $\tau_{L2} = 1$.
\protect\label{fig:fe-vs-tau-ionstrat}}
\end{figure}
\clearpage

Figure \ref{fig:fe-vs-tau-ionstrat} shows $\mathcal{E}$ a tenth
of the way through the wind, or about a continuum mean-free
path from the left edge.
The solid lines show the analytic solution, and
indicate that there are two circumstances in which a globally gray wind
can have a large force efficiency.  Not surprisingly, when
$\tau_{L1} = \tau_{L2}$, the opacity is truly gray,
and $\mathcal{E} = 1$.  What is more surprising is that as
$\tau_{L2}$ is made optically thin, $\mathcal{E}$ also increases,
as the flux becomes more gray even though the opacity does not.
This is because line emissivity dominates,
so the line photons can only be redistributed into other
lines.
Thus as $\tau_{L2}$ decreases, there are fewer and fewer lines
available to receive photons from the thick line region, so
in effect the
reciprocity condition prevents photons from leaving the thick line
region even if that region contains highly redistributing opacity.
Indeed, if $\tau_{L2} = 0$, then $\mathcal{E} = 1$ because no
cross-redistribution can occur.  This is in stark contrast to the SSR model,
which assumes the limit of strong redistribution.

The asterisks in Figure \ref{fig:fe-vs-tau-ionstrat} denote the
locations in parameter space where the photons thermalize, which can be
seen from the exponents of eqs. (\ref{eq:xlsoln}) and
(\ref{eq:xrsoln}) to be when $\tau \approx 1 / \sqrt{a}$.  To the left
of the frequency thermalization point, the photons are not efficiently
redistributed, so little anti-correlation between flux and opacity is
set up, preventing a large drop in $\mathcal{E}$.  This drop does
appear to the right of the asterisk,
where strong frequency thermalization occurs. The crosses
denote the locations in parameter space where $\tau_{L2} = 1$, which is
roughly where the force reaches minimum efficiency.  As
$\tau_{L2}$ gets larger than 1, the gaps fill in and the wind becomes
more gray, allowing a larger $\mathcal{E}$.  Thus $\tau_{L2} = 1$ is
seen as the parameter regime where there is
enough line opacity in the thin-line domain to significantly reduce the
flux in the thick-line domain, but not enough to achieve multiscattering
in the thin-line domain.  Thus we find that multiple momentum deposition
becomes most difficult whenever the line opacity is distributed such
that each ionization zone covers
about a single photon mean-free path for a significant flux-weighted
fraction of the frequency spectrum.
{\it Either} less or more nongrayness in the opacity
will achieve higher overall force
efficiencies.

\subsection{Ionization Stratification With Real Line Lists}
\label{subsec:ionstratreal}

With the schematic two-domain results in mind, we now return to the real
line lists.
We have seen that the radiative flux will respond to the opacity as
though it was effectively gray over
scales much shorter than the frequency thermalization length, which
corresponds to some finite range in
temperature and may include multiple ionization strata.
Over this range, frequency thermalization will take hold and
drive the flux toward a coarse-grained version of the SSR limit, so the
opacity must be appropriately averaged over the ionization states
present.
In this picture, the coarse-grained average SSR opacity controls the local flux,
which then interacts with truly local opacity to determine the radiative
acceleration at each point.

\begin{figure}
\plotone{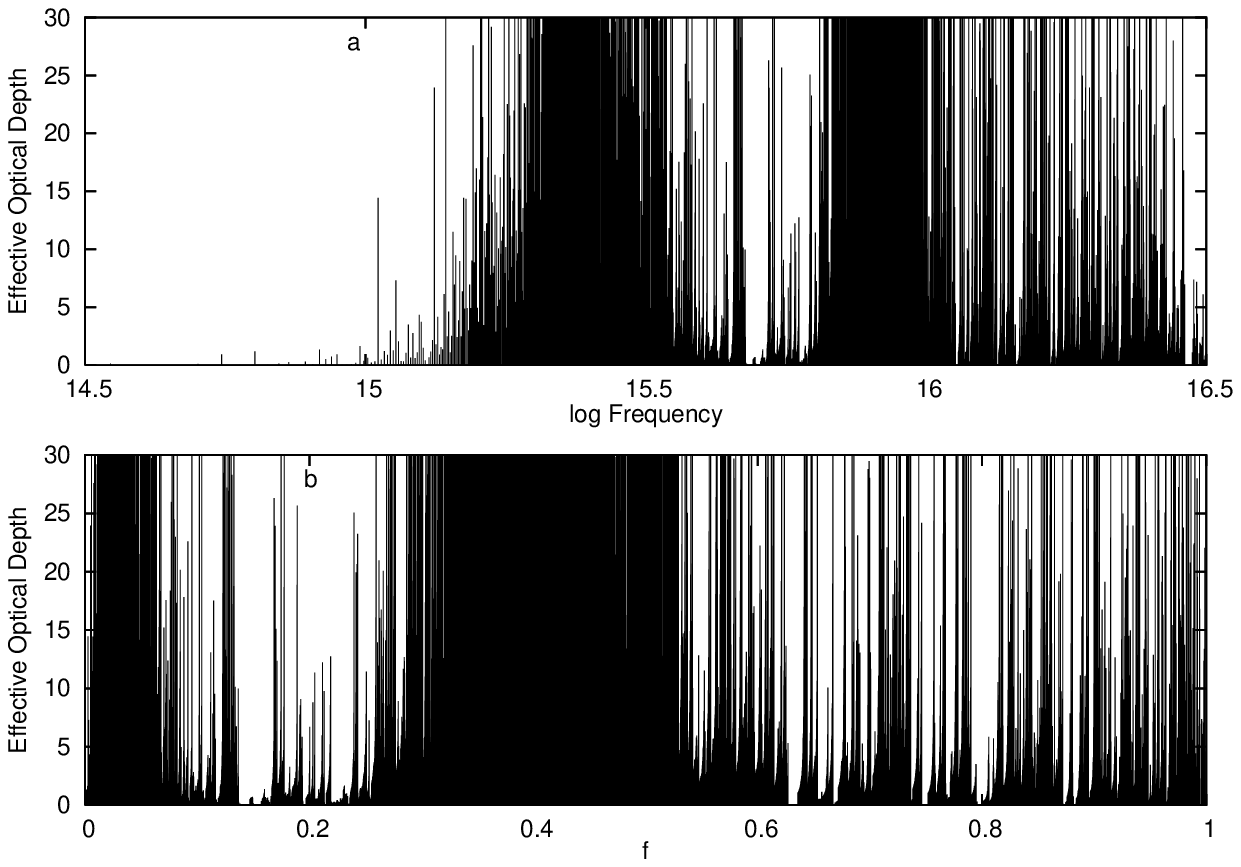}
\caption{The effective optical depth
averaged over the temperature range $8.0\times10^4K \le T \le
1.3\times10^5K$. \protect\label{ld-f-multitemp1}}
\end{figure}

\begin{figure}
\plotone{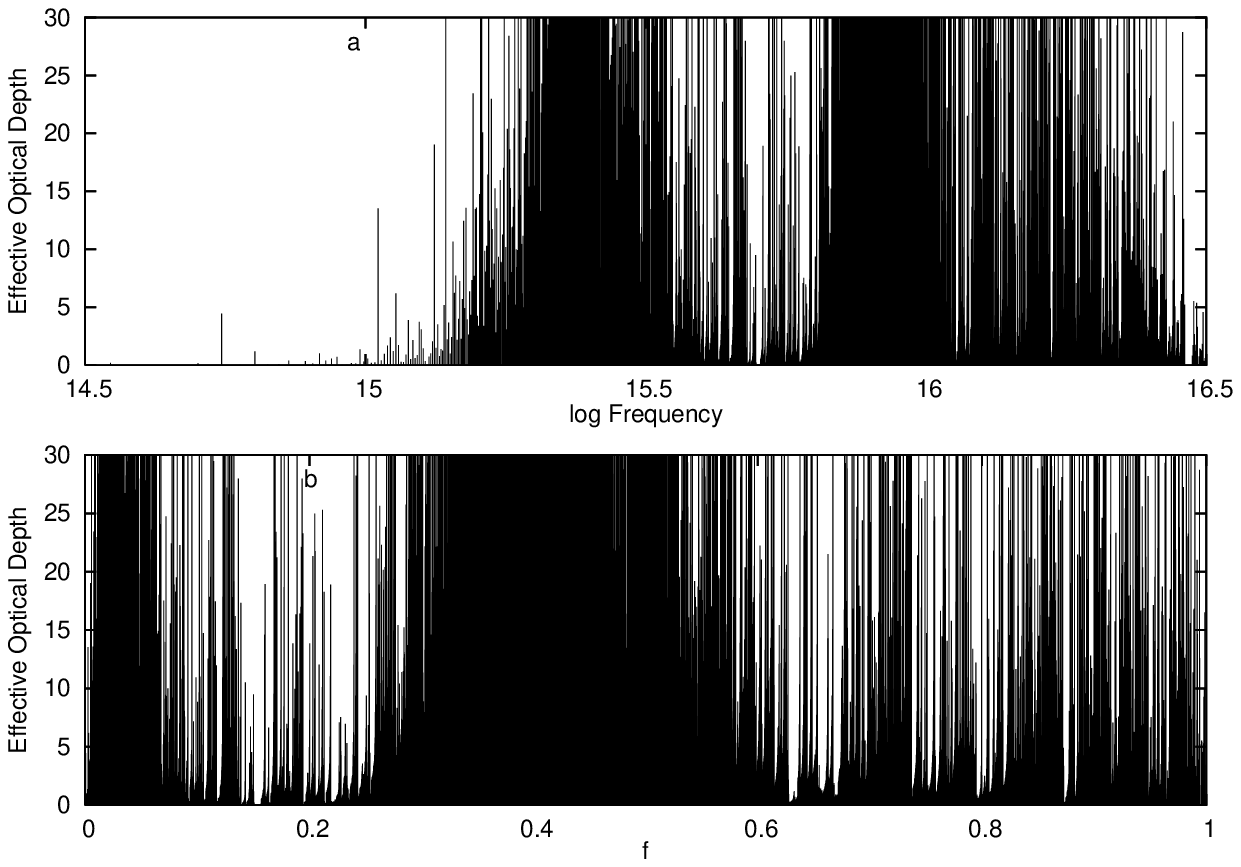}
\caption{The effective optical depth
averaged over the temperature range $6.0\times10^4K \le T \le
1.3\times10^5K$.\protect\label{ld-f-multitemp2}}
\end{figure}

\begin{figure}
\plotone{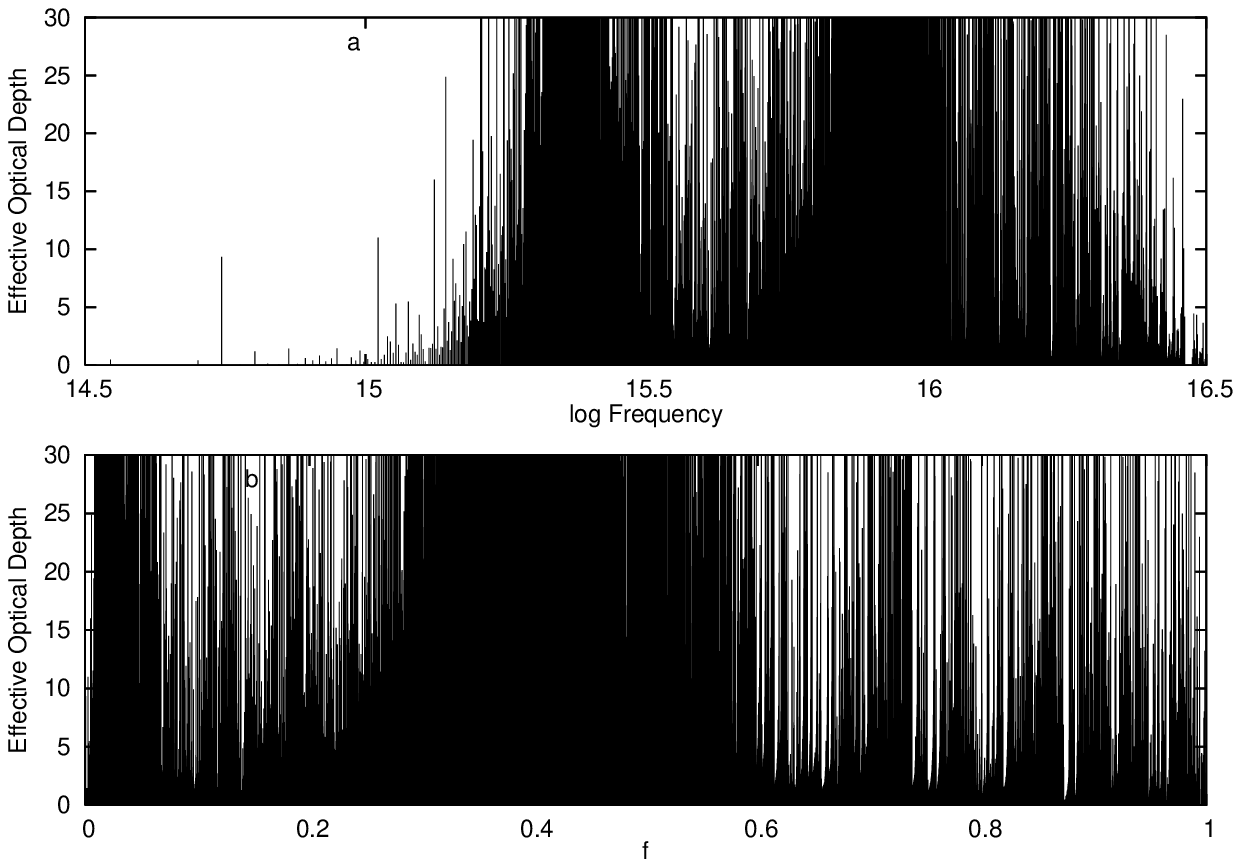}
\caption{The effective optical
depth averaged over the temperature range $4.0\times10^4K \le T \le
1.3\times10^5K$. \protect\label{ld-f-multitemp3}}
\end{figure}

Again, we use the model assumptions in Table \ref{table:assumpt}. The
temperature model of LA93, chosen because of its simplicity and
emphasis on fundamental processes rather than its completeness
relative to more sophisticated models, 
is used to determine the
characteristic maximum range in temperature, some subset of which
would correspond to the frequency thermalization length.  In their
model, the temperature ranges from about $1.35\times10^5K$ at the
stellar surface to about $3.5\times10^4K$ at the free-electron
photosphere.  Without more detailed redistribution calculations (such
as carried out by \citealt{sim2004}), the appropriate temperature range
corresponding to a thermalization depth is unclear, so we consider
temperature ranges of $8.0\times10^4K \le T \le 1.3\times10^5K$,
$6.0\times10^4K \le T \le 1.3\times10^5K$, and $4.0\times10^4K \le T
\le 1.3\times10^5K$ in an effort to span the possibilities. Figures
\ref{ld-f-multitemp1}, \ref{ld-f-multitemp2}, and
\ref{ld-f-multitemp3} show the average effective line optical depths
from the Kurucz list
for these three temperature ranges, respectively.  Notice that the
gaps begin to fill in as the lower temperature limit reaches
$6.0\times10^4K$, and this continues as still lower temperatures are
included. However, we do not encounter contributions from a large
number of ionization states, in seeming contradiction with LA93 but in
agreement with the results of \citet{sim2004}.  For example, the
latter author found that only two stages of iron have a significant
impact on the line-driven mass-loss rate, and this limits the
effectiveness of ionization stratification.

\subsection{Discussion}
\label{ch3iondiscsubsec}

Tables \ref{table:ionstrat} and \ref{table:ionstratop} show the effect
on the force efficiency and mass-loss rate of including a range in
temperatures before applying the SSR flux approximation.  The Kurucz
list results are again insufficient to describe W-R winds.  The OP list
results go as high as $8.9\times10^{-6}\Msun yr^{-1}$.  While such
mass-loss rates have been observed \citep{kurosawa-etal2002}, for this
to be an upper limit on W-R mass-loss would require large amounts of
clumping correction.  This mass-loss rate would also require a flux
themalization length of about 3 stellar radii to
correspond to the required temperature range in the LA93 model, and
this seems unrealistically large given the large potential for
redistribution found by \citet{pinto-eastman2000}.

\clearpage

\clearpage

\clearpage

\clearpage
\begin{deluxetable}{rrrrrrrrr}
\tablecaption{Mass-Loss Rates With Ionization Stratification (Kurucz List)
\protect\label{table:ionstrat}}
\tablewidth{0pt}
\tablehead{\colhead{$\Delta T(K)$} & \colhead{$\alpha$} &
\colhead{$t$} & \colhead{$M(t)$} & \colhead{$\mathcal{E}$} &
\colhead{$a$} &
\colhead{$b$} & \colhead{$y_c$} & \colhead{$\Mdot (\Msun yr^{-1})$}}
\startdata
$1.3\times10^5 - 8.0\times10^4$ & 0.79 & 0.017 & 14.5 & 0.25 & 32 &
5.0 & 1.6 & $4.9\times10^{-6}$\\
$1.3\times10^5 - 6.0\times10^4$ & 0.79 & 0.018 & 13.8 & 0.29 & 50 &
4.6 & 1.8 & $5.7\times10^{-6}$\\
$1.3\times10^5 - 4.0\times10^4$ & 0.80 & 0.020 & 12.7 & 0.34 & 70 &
4.3 & 2.0 & $7.1\times10^{-6}$\\
\enddata
\end{deluxetable}
\clearpage

\begin{deluxetable}{rrrrrrrrr}
\tablecaption{Mass-Loss Rates With Ionization Stratification (OP List)
\protect\label{table:ionstratop}}
\tablewidth{0pt}
\tablehead{\colhead{$\Delta T(K)$} & \colhead{$\alpha$} &
\colhead{$t$} & \colhead{$M(t)$} & \colhead{$\mathcal{E}$} &
\colhead{$a$} &
\colhead{$b$} & \colhead{$y_c$} & \colhead{$\Mdot (\Msun yr^{-1})$}}
\startdata
$1.3\times10^5 - 8.0\times10^4$ & 0.72 & 0.037 & 13.0 & 0.24 & 260 &
5.2 & 1.1 & $7.6\times10^{-6}$\\
$1.3\times10^5 - 6.0\times10^4$ & 0.72 & 0.039 & 12.5 & 0.25 & 260 &
5.1 & 1.2 & $8.3\times10^{-6}$\\
$1.3\times10^5 - 4.0\times10^4$ & 0.72 & 0.041 & 12.0 & 0.26 & 298 &
4.9 & 1.2 & $8.9\times10^{-6}$\\
\enddata
\end{deluxetable}
\clearpage

\section{Conclusions}
\label{sec:concl}

The starting point of this analysis was to recognize that as long as ion
thermal speeds are not artificially enhanced by turbulent processes
\citep[as considered by][]{hamann-grafener2004},
the Sobolev approximation is entirely applicable
to optically thick W-R winds.
Thus CAK theory is applicable after modifying for the effects of
diffuse radiation on the ionization and the angle dependence of
the radiative flux, and it is found that lines are able to drive
abundant mass loss under
hot W-R plasma conditions. However, the
ubiquitous presence of frequency redistribution in an
optically thic flow introduces significant
challenges to achieving large line-driven W-R mass-loss rates, and so
the rate of frequency
thermalization is critical for quantifying how redistribution affects
the efficiency of line driving.

With rapid enough thermalization, the radiative flux avoids large
opacity domains, resulting in force efficiencies well below what is
needed to
drive the observed W-R mass-loss rates.
This drop in force
efficiency in a highly redistributing optically thick
wind may only be avoided by filling the gaps
locally, which requires the discovery of new lines in regions of the
spectrum that are less densely packed.
Our test calculation demonstrated that the greatest challenge to
driving efficiency is
presented by spectral domains in which optically thick lines barely
overlap over the wind terminal speed, 
since regions where the lines are sparse are less likely to
be redistributed into, and regions where the lines are dense are
already effective at line driving.

If, on the other hand, the thermalization rate is slow enough to sample a wide
range in ionization strata, then gray-type force efficiencies may in principle be
recovered.   However, in practice the line lists do not appear to exhibit
sufficiently rich contributions from the many different ionization
states to achieve widespread filling-in of the spectrum.
As a result, frequency redistribution into domains with relatively
little line overlap continues to present a severe challenge to
obtaining line-driven winds with large frequency-averaged optical depth,
as would be required to attain the largest of the Wolf-Rayet
mass-loss rates
inferred from observations.
In short, there still does not exist {\it a priori} opacity-driven models
of supersonically accelerating optically thick winds with low turbulent broadening
and non-static opacity treatments that can
explain how a smaller and hotter W-R star can have a dramatically enhanced mass-loss
rate.
Either the opacity is still incomplete and new lines are capable of filling in the
line-poor domains, or else clumping corrections reduce the need for W-R mass fluxes
to substantially exceed those of their cooler, larger, and H-rich cousins, the extreme
Of stars.

In addition, it should be noted that up-to-date opacity treatments from the
Opacity Project \citep{badnell-seaton2003}
and the OPAL opacity tables \citep{rogers-inglesias1992} are of static type
so are not in their purest form applicable to W-R winds.
They must first be
re-evaluated as expansion-type opacities,  such
as the method of \citet{jeffery1995} or the SSR mean used
here, possibly also incorporating
more exact radiative transfer
\citep{pinto-eastman2000} or non-Sobolev opacity corrections
\citep{wehrse-etal2003},
before they may be appropriately applied to further investigations into
the role of high-temperature opacity contributions (such as the ``iron bump'')
in explaining high W-R mass fluxes.
Furthermore, the critical point must not be artificially placed in regions of
exceptionally high opacity, as this would belie the meaning of the critical
point as the ``choke point'' of wind acceleration.`

It must also be mentioned that although clumping corrections reduce both
the inferred W-R mass-loss rates and the difficulty in explaining them
with existing line opacities, clumping itself may introduce
dynamical challenges.  This is not a problem for clumps smaller than the
Sobolev length $L$, but larger clumps will reduce the force
efficiency in ways that are classifiable according to whether the
clumps are optically thin or thick to most photons. When the clumps are
thin, their impact is felt only through the role of density and velocity
in standard CAK theory, but when the clumps are thick
\citep[e.g.,][]{brown-etal2004}, additional reductions in the force
efficiency must appear owing to the feedback onto the self-consistent
radiative flux, in a manner again similar to the spirit of a Rosseland
mean \citep{owocki-etal2004}. The development and dynamical
implications of such clumps require detailed radiation hydrodynamic
simulations, but the simplified approaches developed here may be used
to guide approximations that make such a time-dependent calculation
\citep[e.g.,][]{baron-hauschildt2004, grafener-hamann2005} computationally tractable.

Finally, the most important goal of this paper has been to develop a conceptual
vocabulary to discuss the circumstances under which a W-R wind may or
may not be efficiently driven by Sobolev-type line opacity.
Key elements of this vocabulary include the effectively gray optical
depth  $\tau_g$, the nongray force efficiency
$\mathrel{E}$, the nongrayness parameter $b$ for the monotonically
reordered line distribution, and the range of ionization strata that
contribute within a photon frequency thermalization length.
Estimations of these parameters offer conceptual insights into
classifying various physical behaviors, both before and after carrying
out
optically thick radiation hydrodynamical simulations.

\acknowledgments{The authors would like to thank John Bjorkman and
  Ivaylo Mihaylov for code contributions and discussion, and John
  Hillier for insightful comments.  This project was supported by the
  National Science Foundation (AST 00-98155). Portions of this work
  were performed under the auspices of the U.S. Department of Energy
  by Los Alamos National Laboratory under contract No. W-7405-ENG-36}

\end{document}